\documentclass[12pt]{spieman}  
\usepackage{amsmath,amsfonts,amssymb}
\usepackage{graphicx}
\usepackage{setspace}
\usepackage{tocloft}
\usepackage{lineno}
\title{Far-infrared probing with PRIMA 
    into particle acceleration associated with relativistic jets 
    from active galactic nuclei}
\author[a,b,*]{Naoki Isobe}
\author[c,a]{Takao Nakagawa}
\author[d]{Motoki Kino}
\author[e,f]{Hiroshi Nagai}
\author[a]{Shunsuke Baba}
\author[g]{Makoto Tashiro}
\affil[a]{
    Institute of Space and Astronautical Science (ISAS), 
    Japan Aerospace Exploration Agency (JAXA), 
    3-1-1 Yoshinodai, Chuo-ku, Sagamihara, Kanagawa, 252-5210, Japan}
\affil[b]{
    Kwansei Gakuin University, 1 
    Uegahara, Gakuen, Sanda, Hyogo, 669-1330, Japan}
\affil[c]{
Advanced Research Laboratories, Tokyo City University, 1-18-1, 
Tamazutsumi, Setagaya, Tokyo, 158-8557, Japan 
}
\affil[d]{
    Kogakuin University of Technology \& Engineering, 
    Academic Support Center,
    2665-1 Nakano, Hachioji, Tokyo, 192-0015, Japan}
\affil[e]{
    National Astronomical Observatory of Japan,
    2-21-1 Osawa, Mitaka, Tokyo, 181-8588, Japan}
\affil[f]{
    Department of Astronomical Science, 
    The Graduate University for Advanced Studies, 
    SOKENDAI, 2-21-1 Osawa, Mitaka, Tokyo 181-8588, Japan}
\affil[g]{
    Department of Physics, Saitama University, 
    255 Shimo-Okubo, Sakura-ku, Saitama, 338-8570, Japan}

\cftpagenumbersoff{figure}
\cftpagenumbersoff{table} 
\begin{document} 
\maketitle
\begin{abstract}
It is presented that 
the Probe far-Infrared Mission for Astrophysics (PRIMA) 
has a high potential to study particle acceleration phenomena 
associated with jets emanating from active galactic nuclei.
A special focus is put on hot spots of radio galaxies
because they are widely regarded as the jet-terminal shock
where particles are accelerated via the diffusive shock acceleration.
To investigate the particle acceleration condition 
in the hot spots, 
it is of prime importance to evaluate their magnetic field strength. 
As a useful indicator of the magnetic field, 
we propose to adopt a synchrotron spectral feature called the cooling break,
of which the frequency is determined by the mutual balance 
between the synchrotron radiative cooling and the adiabatic one. 
Referring to the standard physical parameter of the hot spots, 
the cooling break is expected to reside 
in or slightly below the far-infrared range covered with PRIMA.
The feasibility of the PRIMA observations 
to measure the far-infrared flux density and 
to constrain their cooling break frequency 
is discussed for nearby well-studied hot spots.
An affordable observational strategy with PRIMA is described. 
A possible application of the method to lobes of radio galaxies 
is also briefly discussed.
\end{abstract}

\keywords{far infrared;
    astronomy;
    photometry;
    infrared imaging;
    synchrotron radiation}

{\noindent \footnotesize\textbf{*}Naoki Isobe,  \linkable{n-isobe@ir.isas.jaxa.jp} }

\begin{spacing}{2}   

\section{Introduction} 
\label{sec:intro}   
Astrophysical relativistic jets emanating 
from active galactic nuclei are regarded as
one of the largest and most energetic accelerators 
in the Universe.
In particular, compact hot spots of 
Fanaroff-Riley class-II (FR-II) radio galaxies \cite{Fanaroff1974}
with a typical physical scale of a few kpc
are widely associated with their jet-terminal shocks,
where particles are energized 
through the diffusive shock acceleration, 
or the so-called Fermi-I process \cite{Bell1978,Begelman1984}. 
The particles accelerated in the hot spots are thought to 
diffuse out nearly adiabatically into the intergalactic space, 
forming radio lobes extending namely on a $\gtrsim 100$ kpc scale.
Thus, these jet-related structures are proposed 
as one of the promising candidates 
for the origin of highenergy cosmic rays 
\cite{Meisenheimer1989,Meisenheimer1997}.
Actually, anisotropy in the arrival directions 
of ultra-high-energy cosmic rays 
suggested by observations performed with the Pierre Auger Observatory \cite{Aab2018}
is proposed to be ascribed to the radio galaxies \cite{Matthews2018}.

In the process of particle acceleration,
magnetic fields are theoretically thought to play 
several fundamental roles \cite{hillas1984, kotera2011}.
These include the assistance of energy transfer 
from the plasma flow to individual particles,
particle confinement within the acceleration area,
radiative cooling experienced by the accelerated particles, and so forth.
Therefore, 
it is of prime importance to evaluate the magnetic field, $B$, 
to constrain the particle acceleration condition 
in the hot spots and lobes of radio galaxies. 

Conventionally, over the last two to three decades,
a comparison between the synchrotron radio 
and inverse Compton (IC) X-ray intensities 
has been a standard method to estimate 
the magnetic field strength in the hot spots and lobes.
The synchrotron radio and IC X-ray flux densities, respectively, 
scales as $S_{\rm R} \propto U_{\rm e} U_{\rm B}$ and 
$S_{\rm X} \propto U_{\rm e} U_{\rm seed}$,
where $U_{\rm e}$, $U_{\rm B}$ ($\propto B^2$) and $U_{\rm seed}$ denote,
respectively, the energy densities of electrons, 
magnetic fields and seed photons. 
As far as the seed photon source is specified, 
the radio-to-X-ray flux-density ratio is regarded as 
a good indicator of the magnetic field 
as $U_{\rm B} \propto S_{\rm R}/S_{\rm X}$,  
on a simple assumption, e.g.,
that the electron number density spectrum 
and corresponding synchrotron one are described 
by a single power-law (PL) model.
In the case of the hot spots, the seed photons are typically 
dominated by the synchrotron radiation itself 
(the so-called synchrotron-self-Compton mechanism 
or the SSC one \cite{Band1985}), 
whereas in the lobes, the cosmic microwave background radiation 
frequently acts as the dominant seed photon source \cite{Harris1979}.
This method was widely applied 
to well-studied hot spots \cite{Hardcastle2004,kataoka2005} 
and lobes \cite{Kaneda1995,Isobe2002,Isobe2005,Isobe2009}, 
from which the IC X-ray flux density was measured 
with X-ray observatories including ASCA, Chandra, XMM-Newton, and Suzaku.
Through these studies, 
the equipartition condition between the electrons and magnetic fields
(i.e., $U_{\rm e} \simeq U_{\rm B}$) was systematically  
tested \cite{Hardcastle2004, kataoka2005},
and a significant deviation toward the electron dominance 
(i.e., $U_{\rm e} \gg U_{\rm B}$)
has been suggested in some objects \cite{Isobe2002,Hardcastle2004}. 

To be precise, however, 
the above magnetic field evaluation is too simplified. 
Theoretical studies \cite{Carilli1991} widely predict that 
the standard diffusive shock acceleration 
under the continuous energy injection condition 
generates a broken PL-like electron energy distribution \cite{Carilli1991},
and thus, 
the corresponding synchrotron spectrum is also assumed 
to be described with the broken PL form.
Thus, the magnetic field estimated from the simple flux ratio,
$S_{\rm R}/S_{\rm X}$, alone is inevitably subjected to uncertainties 
originating from the assumption 
of the synchrotron radio and IC X-ray spectral shape.

The break of the electron and synchrotron spectra, instead,
is regarded as an independent indicator of the magnetic field
because its frequency is determined by a mutual balance 
between an electron radiative cooling 
and an adiabatic loss \cite{Inoue1996}.
Thus, the break is widely called the cooling break. 
As the radiative cooling in the hot spots is 
typically dominated by synchrotron radiation \cite{Kino2004},
the magnetic field is estimated from the cooling break frequency.
The advantage of the method is that 
it basically independent of whether the IC X-ray emission is detected or not. 

An increasing use of submillimeter and/or infrared data 
for the hot spots 
\cite{Isobe2017, Isobe2020, Sunada2022, Werner2012}
has gradually initiated to grasp the observational evidence of the cooling break.
From mid-infrared data obtained with the Spitzer observatory,
the well-studied hot spots are implied to exhibit 
a spectral break in the range of 
$\nu_{\rm b} = 10^{11}$--$10^{13}$ Hz \cite{Werner2012}.
Therefore, 
a combination of radio, submillimeter and far-infrared observations 
is useful to constrain the break frequency. 
Actually, this technique was successfully applied to 
the hot spot D of the nearby prototypal FR-II radio galaxy Cygnus A \cite{Sunada2022},
by utilizing the far-infrared data taken 
with the Herschel observatory.
The study indicates that 
far-infrared data are of crucial importance 
to measure the cooling break frequency and,
thus, to evaluate the magnetic field.
However, the limited sensitivity of Herschel 
prevented a systematic magnetic field evaluation 
in the hot spots of radio galaxies. 

The PRobe far-Infrared Mission for Astrophysics (PRIMA) \cite{Glenn2025} 
is a cryogenically cooled far-infrared observatory concept, 
which has been selected as a candidate 
for the NASA Probe Explore mission, with the target launch in 2032.
Among the two science instruments onboard PRIMA,
PRIMAger\cite{Ciesla2025} offers two-types 
of high-performance far-infrared imagers.
The PRIMA Hyperspectral Imager (PHI) operates
in the wavelength range of $\lambda = 24$--$84$ $\mu$m 
(or the corresponding frequency range of $\nu = (12.5$--$3.6)\times 10^{12}$ Hz)
with a spectral resolution of $R=10$,
whereas the PRIMA Polarimetric Imager (PPI) 
enables four-band polarimetric imaging in the range of $\lambda = 80$--$261$ $\mu$m 
($\nu = (3.7$--$1.1)\times10^{12}$ Hz). 
Coupled with the $4.5$ K cryogenic telescope 
with an aperture diameter of $1.8$ m,
PRIMAger ensures an unprecedented far-infrared sensitivity, 
which is by more than an order of magnitude better 
than previous instruments in the similar wavelength range such 
as the Herschel observatory.
These properties make PRIMAger the ideal instrument 
to systematically detect the far-infrared emissions from the hot spots
and to measure their magnetic field through the cooling break.

\section{Method} 
\label{sec:cooling_break}
The diffusive shock acceleration 
under the continuous energy injection condition \cite{Carilli1991}
is assumed as the standard acceleration condition in the hot spots.
The electron Lorentz factor at the cooling break is determined by 
the mutual balance between the radiative cooling timescale of the electrons
and the adiabatic or dynamical timescale of the plasma flow 
in the post-shock region \cite{Inoue1996}.
Because the radiative cooling in the hot spots is suggested 
to be typically dominated by the synchrotron radiation \cite{Kino2004}, 
the radiative cooling timescale is described as 
$\tau_{\rm cool} 
    = \frac{3 m_{\rm e} c}{4 U_{\rm B} \sigma_{T} \gamma_{\rm e}} 
    = \frac{6 \pi m_{\rm e} c}{B^2 \sigma_{T} \gamma_{\rm e}}$
in the cgs unit system, 
where $m_{\rm e}$, $c$, $\sigma_{\rm T}$ and $\gamma_{\rm e}$
are the electron mass, speed of light, Thomson cross-section, 
and electron Lorentz factor, respectively. 
The adiabatic timescale is denoted as $\tau_{\rm ad} = \frac{L}{v}$,
with $L$ and $v$, respectively, 
being the source size along the flow 
and the flow speed in the downstream region of the shock 
evaluated in the shock frame.
By equating these two timescales as $\tau_{\rm cool} = \tau_{\rm ad}$, 
the electron Lorentz factor at the cooling break is given 
as 
\begin{equation} 
    \gamma_{\rm b} 
    = \frac{6 \pi m_{\rm e} v c}{B^2 \sigma_{\rm T} L_{\rm cool}}.
\label{eq:break_gamma}
\end{equation} 
Here and hereafter, 
the source size $L$ corresponding to the cooling break 
is re-defined as the cooling length $L_{\rm cool}$
because the radiative cooling becomes effective 
after the plasma travels the distance of $L_{\rm cool}$.

Based on the equation for the synchrotron critical frequency,
$\nu_{\rm c} \simeq \frac{3 \gamma_{\rm e}^2 e B}{4 \pi m_{\rm e} c}$,
the break Lorentz factor in Eq. (\ref{eq:break_gamma}) 
is converted into the break synchrotron frequency as
\begin{equation} 
    \nu_{\rm b} = 
    \frac{27 \pi e m_{\rm e} v^2 c}
        {\sigma_{\rm T}^2} B^{-3} L_{\rm cool}^{-2}.  
\label{eq:break_freq}
\end{equation} 
Equation (\ref{eq:break_freq}) indicates 
that the break frequency in the synchrotron spectrum 
is sensitive to both $B$ and $L_{\rm cool}$.
Because the cooling length $L_{\rm cool}$ (i.e., the source size) 
is usually measurable from high-resolution interferometric radio images,
an observational constraint on the break frequency yields 
a reliable estimate of the magnetic field strength. 
This method has been successfully applied to a few hot spots
by making use of far-infrared data obtained 
with the Herschel observatory 
\cite{Isobe2017,Isobe2020,Sunada2022,Isobe2023}.

\section{Feasibility} 
\label{sec:feasibility}
\subsection{Preliminary Investigation} 
\label{sec:prelim_investigatoin}
\begin{figure}[t]
\centering
\includegraphics[width=0.7\textwidth]{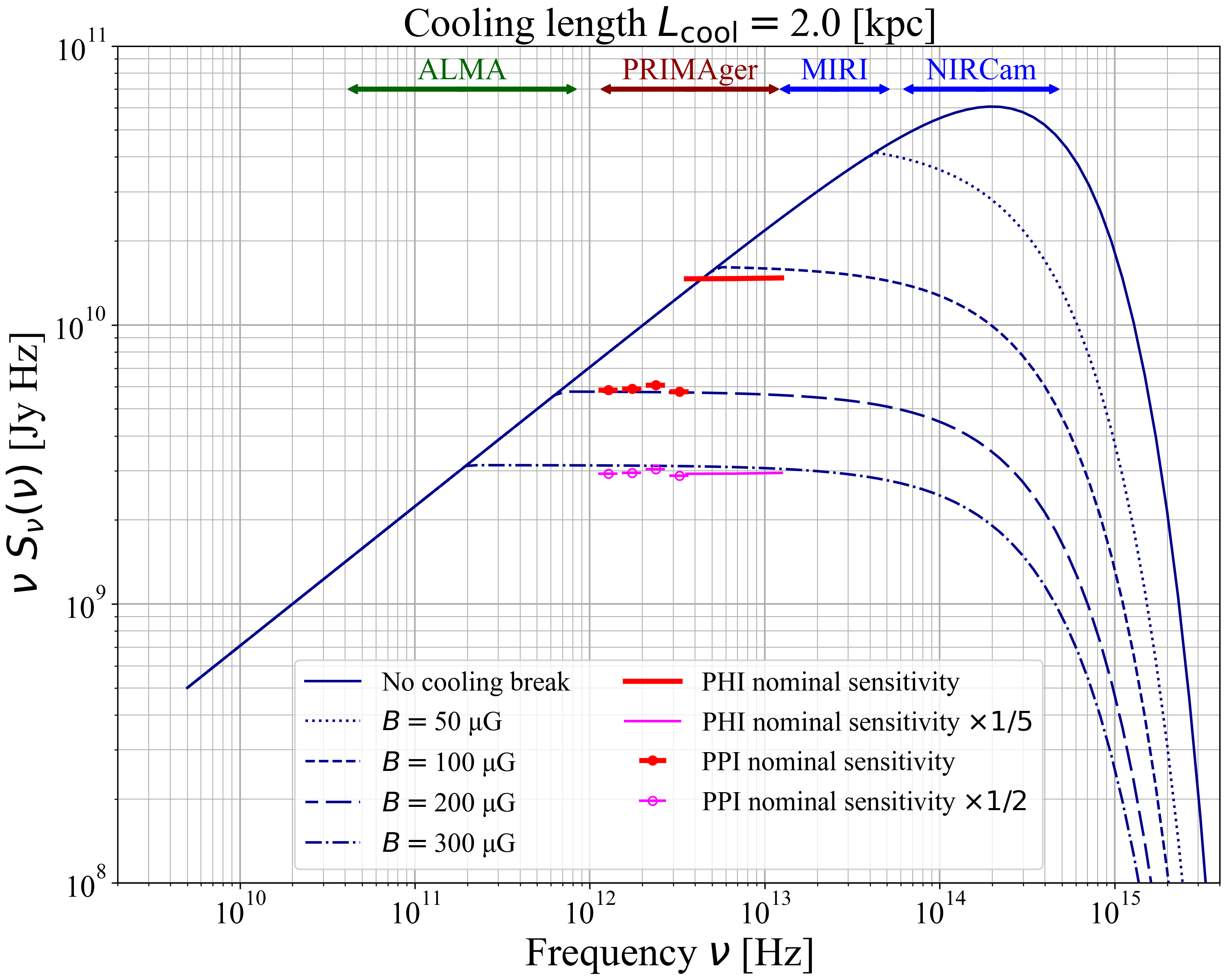}
\caption{Magnetic-field dependence 
of the synchrotron spectral energy distribution of the hot spots.
The magnetic field of $B = 50$, $100$, $200$ and $300$ $\mu$G
is adopted.
The break frequency is estimated from Eq. (\ref{eq:break_freq})
by assuming the cooling length of $L_{\rm cool} = 2$ kpc,
where the flow speed of $v = 0.3 c$ is employed.
The spectral cut-off 
at the frequency of $\nu_{\rm c} = 4\times10^{14}$ Hz is implemented.
The spectral normalization of $S_\nu(5~{\rm GHz}) =0.1$ Jy is employed.
The synchrotron spectrum unaffected by the radiative cooling 
is also shown for comparison.
The estimated spectra are compared with the PPI and PHI nominal sensitivities.
The horizontal arrows at the top indicate the spectral coverages
with ALMA, PRIMAger, JWST MIRI and NIRCam. 
} 
\label{fig:sed}
\end{figure}

Figure \ref{fig:sed} displays 
the variation of the synchrotron spectral energy distribution 
as a function of the magnetic field strength $B$.
In the plot, as representative values,
the magnetic field strength of
$B=50$, $100$, $200$ and $300$ $\mu$G is adopted. 
A broken PL model subjected 
to a high-frequency spectral cut off is simply adopted, 
instead of performing a detailed synchrotron calculation.
The cut-off frequency is fixed at $\nu_{\rm c} = 4 \times10^{14}$ Hz,
a typical value for the hot spots 
with a good-quality mid-infrared data \cite{Werner2012}. 
The spectrum is normalized to the flux density of $S_{\nu} = 0.1$ Jy 
at the frequency $\nu = 5$ GHz,
which is the median value of the well-studied X-ray-detected hot spots
listed in Ref. \citenum{Hardcastle2004}.
The cooling length of $L_{\rm cool} = 2$ kpc is employed
because this corresponds to the median size of the hot spots  
tabulated in Ref. \citenum{Hardcastle2004}.
The break frequency is derived from the adopted $B$ and $L_{\rm cool}$ 
values through Eq. (\ref{eq:break_freq}).
It is widely thought that jets in FR-II radio galaxies are 
relativistic even at the position of their hot spots.
Theoretical studies predict a downstream flow velocity of $v=\frac{1}{3} c$ 
for highly relativistic shocks\cite{Kirk1999,Kino2004}.
In the following, the down-stream velocity of $v=0.3c$ 
(nearly corresponding to the highly relativistic shock)
is employed as a reference value
by referring to previous mid-to-far infrared studies 
of hot spots \cite{Isobe2017,Isobe2020,Isobe2023}.
Following the strong shock condition, 
the spectral index below the break frequency of $\alpha_1 = 0.5$ is adopted,
where the index is defined as $S_\nu \propto \nu^{-\alpha}$.
Assuming the standard cooling break 
under the continuous energy injection condition, 
the index change at the break is 
set at $\Delta \alpha = 0.5$ \cite{Carilli1991}, and thus,
the higher frequency index becomes $\alpha_2 = 1.0$.
In Fig. \ref{fig:sed}, 
a synchrotron spectrum without the cooling break is also plotted 
as a comparison. 
The $5\sigma$ PRIMAger sensitivity   
to be obtained in the nominal one-square-degree survey 
for a total duration of $10$ h \cite{Ciesla2025}
(as of 2025 February) is displayed in Fig. \ref{fig:sed}.

\begin{figure}[t]
\centering
\includegraphics[width=0.95\textwidth]{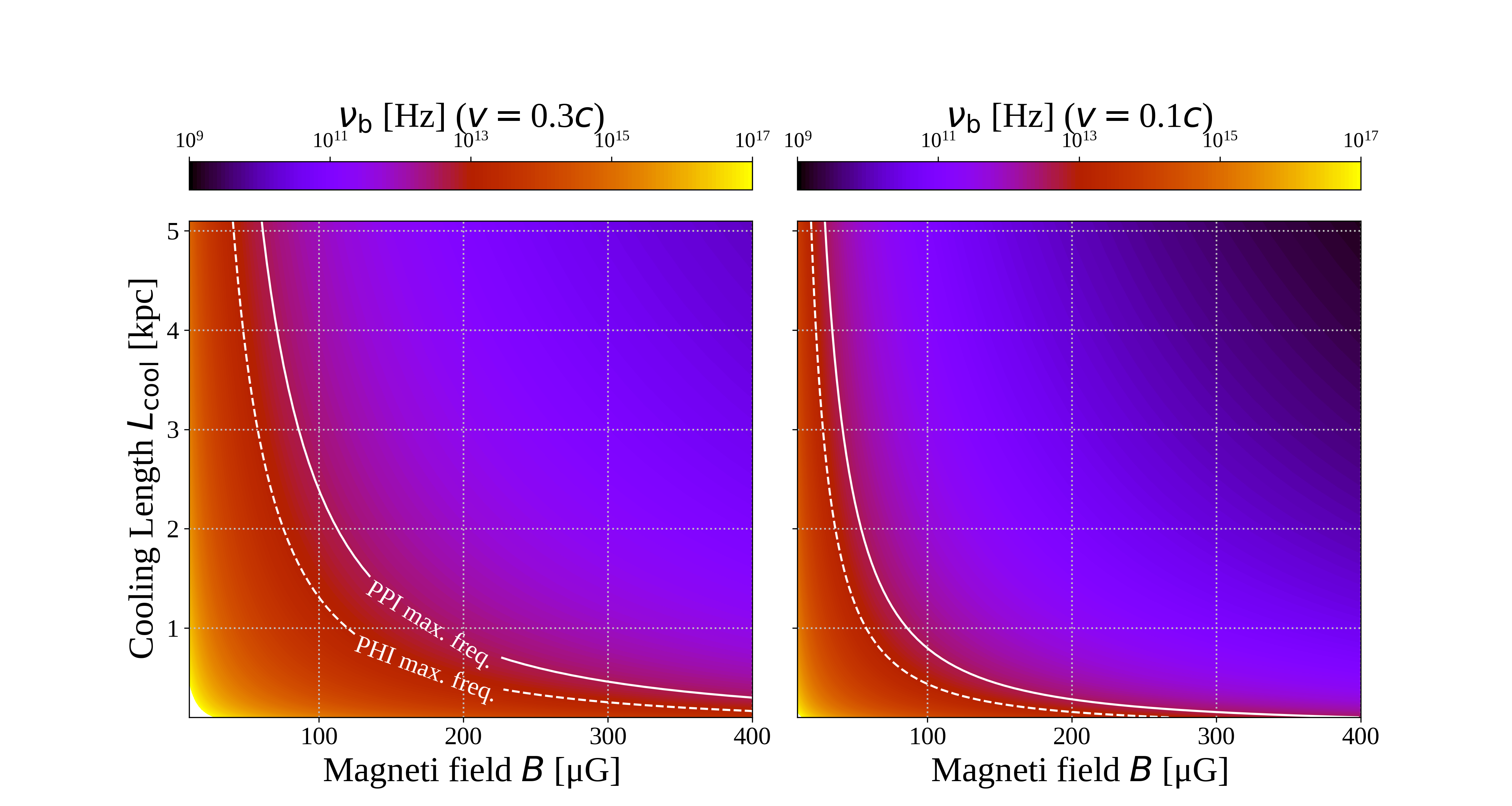}
\caption{Cooling break frequency $\nu_{\rm b}$
for the flow velocity of $v=0.3c$ (left panel) and $0.1c$ (right panel),
plotted as a function of the magnetic field $B$ 
and cooling length $L_{\rm cool}$.
The solid and dashed lines show the maximum frequencies 
covered by the PPI ($3.7 \times 10^{12}$ Hz) and 
PHI ($1.25 \times 10^{13}$ Hz), respectively.} 
\label{fig:break_map}
\end{figure}

First, the PRIMAger spectral coverage is briefly compared 
with the cooling break frequency predicted 
for hot spots with typical physical parameters.
It is reported that the magnetic field strength for the well-studied hot spots,
derived from the SSC modeling to the radio and X-ray spectra,
is typically distributed in the range of $B=100$--$300$ $\mu$ G \cite{kataoka2005}.
Figure \ref{fig:sed} indicates that 
in combination with the radio data in the GHz range,
the PRIMAger data are expected to be useful to constrain 
the cooling break for the magnetic field in this range,
in the case of the ``median'' hot spot with $L_{\rm cool}= 2$ kpc.
The power of PRIMAger for detecting $\nu_{\rm b}$ 
is more clearly visualized in Fig. \ref{fig:break_map},
which presents the $\nu_{\rm b}$ map on the $B$--$L_{\rm cool}$ plane.
The left panel assumes the reference flow velocity, i.e., $v=0.3c$. 
The maximum frequency covered by the PPI 
($\nu = 3.7 \times 10^{12}$ Hz or $\lambda = 80$ $\mu$m)
is drawn with the solid line on the $\nu_{\rm b}$ map.
The PPI, in combination with auxiliary radio observatories,  
is expected to be applicable to hot spots 
of which the $B$ and $L_{\rm cool}$ values are located 
in the upper right of this line.
With the PHI, with the spectral coverage that reaches 
up to $\nu = 1.25 \times 10^{13}$ Hz ($\lambda = 24$ $\mu$m),
the observable parameter space will be enlarged to the dashed line,
although its sensitivity in the nominal survey observation 
is slightly low.
Figure \ref{sec:cooling_break} ensures that 
PRIMAger widely covers the well-studied hot spots
because their size is typically distributed 
in the range of $0.3$ to $10$ kpc \cite{Hardcastle2004}.

Equation (\ref{eq:break_freq}) indicates that 
the cooling break exhibits a relatively strong dependence 
on the post-shock flow velocity as $\nu_{\rm b} \propto v^2$,
although the $v$ value has remained 
yet observationally well unknown. 
To visualize the $v$ dependence, the $\nu_{\rm b}$ map for $v=0.1 c$ 
(corresponding to a mildly relativistic jet)
is plotted in the right panel of Fig. \ref{fig:break_map}.
In comparison to the reference case ($v=0.3c$), 
the break frequency is found to be reduced by a factor of $9$
for $v=0.1c$.
Thus, it is suggested that as the flow gets slower,
the PRIMAger coverage is shifted to weaker magnetic field
and/or shorter cooling-length objects. 
A possible method to constrain the downstream flow velocity 
is briefly discussed in Sec. \ref{sec:synergy_IC}.

In the next step, a crude investigation is performed 
into whether the PRIMAger sensitivity is sufficient 
to detect the synchrotron emission from the hot spots.
Throughout the far-infrared flux-density evaluation and related investigation,
the cooling length and flow velocity are commonly assumed 
as $L_{\rm cool}=2$ kpc and $v=0.3c$, respectively.
Figure \ref{fig:sed} indicates that 
the cooling break frequency is predicted to be lower,
and the corresponding far-infrared synchrotron flux density 
tends to be lower 
as the higher magnetic field value is adopted for the fixed radio flux density.
Hence, the magnetic field strength of $B = 300$ $\mu$G is 
adopted here as the representative upper-end value 
of the well-studied hot spots \cite{kataoka2005}
because the value is thought to predict 
a reasonable lower limit on the far-infrared flux for individual objects.
Correspondingly, the dash-dotted line in Fig. \ref{fig:sed}
is utilized as the spectral template.
The far-infrared flux density at the individual PRIMAger channels
for the radio flux density $S_\nu ({\rm 5~GHz}) = 0.1$ Jy 
is estimated as listed in Table \ref{tab:flux_estimation}. 
To detect the estimated flux density, 
the nominal PPI and PHI survey sensitivities 
are required to be improved by a factor of $2$ and $5$ respectively, 
as shown in Fig. \ref{fig:sed}.
Thus, it is  suggested that by optimizing 
the survey area and total duration within a reasonable manner, 
both PPI and PHI become possibly applicable 
to hot spots in a considerable flux range, 
as detailed in Sec. \ref{sec:strategy}.

\begin{table}[t]
    \centering
    \caption{Synchrotron flux-density estimation in the PRIMAger range 
    for the input physical values of 
    $L_{\rm cool}=2$ kpc, 
    $S_\nu (5~{\rm GHz})=0.1$ Jy, 
    $v=0.3c$, and $B=300$ $\mu$G.}
    \label{tab:flux_estimation}
    \begin{tabular}{l|c|c|c|c}
    \hline
    Channels& $\lambda$~($\mu$m)
            & $\nu$ ($10^{12}$ Hz)
            & $S_\nu$~(mJy)
            & $t_{5'\times5'}$~(h)\\
    \hline
    PPI4    & $235$ & $1.28$ & $2.46$  & $0.24$     \\
    PPI3    & $172$ & $1.74$ & $1.80$  & $0.25$     \\
    PPI2    & $126$ & $2.38$ & $1.31$  & $0.26$     \\
    PPI1    & $ 92$ & $3.26$ & $0.96$  & $0.24$     \\
    \hline
    PHI2    & $ 84$ & $3.57$ & $0.87$  & $1.53$     \\
    PHI2/1  & $ 45$ & $6.66$ & $0.46$  & $1.56$     \\
    PHI1    & $ 24$ & $12.5$ & $0.24$  & $1.62$                   \\
    \hline            
    \end{tabular} \\
\end{table}

As the nucleus of FR-II radio galaxies 
tends to be brighter than their hot spots 
in the far-infrared range\cite{Isobe2020},
it is important to take care of the nuclear contamination 
onto the hot spots for the PRIMAger photometry.
Thus, target sources for the present study 
are possibly limited by the PRIMAger beam size;
4.1 arcsec in the Full Width at Half Maximum (FWHM) 
for the PHI1 channel 
and 10.8 arcsec for the PPI1\cite{Ciesla2025} one.
Conservatively, for a hot spot to be 
free from the far-infrared emission from the nucleus, 
their angular separation is requested to be larger than 
$\sim 10$ and $\sim 30$ arcsec for the PHI and PPI, 
respectively (corresponding to $\sim 3$ times the FWHM beam size). 
For hot spots with a smaller separation, 
a careful subtraction of the nuclear far-infrared emission 
is necessary. 

The PRIMAger beam size is expected to be generally larger than
that of other instruments, especially radio interferometers.
Thus, simple aperture photometry probably suffers 
from systematic uncertainties due to the large PRIMAger beam.
Instead, to precisely measure the PRIMAger flux density 
of the target hot spot,
it is required to fit the observed image with its source size,
which is convolved with the PRIMAger beam.
The beam-convolved image fitting method is also useful for subtracting 
the far-infrared emission from contaminating sources (including the nucleus).

\begin{figure}[t]
    \centering
    \includegraphics[width=0.95\linewidth]{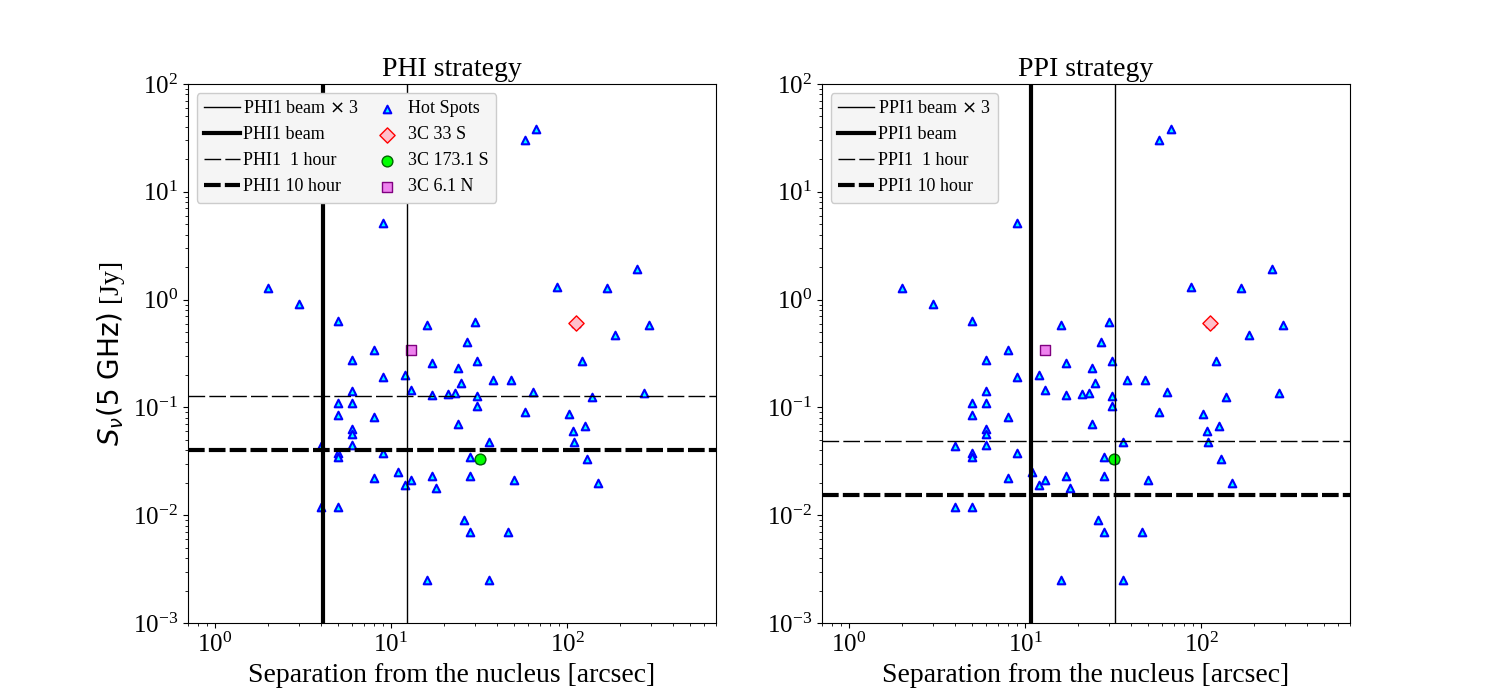}
    \caption{
    Relation between the radio flux density $S_\nu (5~{\rm GHz})$
    and the angular separation from the nucleus 
    for the candidate hot spots 
    \cite{Hardcastle2004,kataoka2005,Meisenheimer1997,
            Mack2009,Cheung2005,Werner2012} (the triangles).
    The representative hot spots picked up in Sec. \ref{sec:candidates} 
    are indicated in both panels with 
    the diamond (the south hot spot of 3C 33), 
    circle (the south hot spot of 3C 173.1) 
    and box (the north hot spot of 3C 6.1).
    The radio intensity, corresponding to the PHI1 sensitivity 
    to be obtained  in the $5'\times5'$ mapping 
    for the exposure time of 1 and 10 h, is drawn 
    with the thin and thick horizontal dashed lines, respectively,
    in the left panel.     
    The thick and thin vertical solid lines in the left panel display 
    the FWHM PHI1 beam size and three times that, respectively.     
    Those for the PPI1 channel are shown in the right panel.
    } 
    \label{fig:strategy}
\end{figure}

\subsection{Observation Strategy} 
\label{sec:strategy}
By referring to the basic feasibility study presented 
in Sec. \ref{sec:prelim_investigatoin},
a more realistic observational plan is proposed. 
For this purpose, a candidate target sample was constructed 
by compiling well-studied hot spots picked up from the literature
\cite{Hardcastle2004,Werner2012,Mack2009,Meisenheimer1997,
Cheung2005,kataoka2005}.
Because there are basically sufficient radio data 
to evaluate the low-frequency spectral index  
for all the selected objects,
a far-infrared detection with PRIMAger  
inevitably enables evaluation of their cooling break frequency 
$\nu_{\rm b}$ and hence the magnetic field strength $B$. 
The separation between the hot spot and nucleus in the sample
is found to be distributed in the range of $\lesssim 4.9$ arcmin.
To simultaneously observe the hot spot and nucleus,
a mapping strategy with a narrow field coverage 
of $5$ arcmin $\times 5$ arcmin 
(hereafter referred to as the $5'\times 5'$ mapping) 
is commonly applied to all the listed hot spots, here, for simplicity. 
This mapping area is comparable to the field of view of each PPI channel.
Therefore, to obtain multiband PRIMAger data,
independent mappings are required 
for the individual PHI and/or PPI channels. 
The exposure time necessary to detect an object 
with the far-infrared flux density shown in Table \ref{tab:flux_estimation}
is estimated by the PRIMA Exposure Time Calculator 
(\linkable{https://prima.ipac.caltech.edu/page/etc-calc}) 
as of 2025 February.
The result is tabulated as $t_{5'\times5'}$ in Table \ref{tab:flux_estimation}.
The estimated exposure time with the $5'\times5'$ mapping 
is regarded as reasonable for all the PHI and PPI channels. 

Figure \ref{fig:strategy} compares the observational characteristics 
of the candidate hot spots (the triangles) 
with the PRIMAger instrumental performance 
on the plane of the radio flux, $S_\nu (5~{\rm GHz})$, 
and nuclear separation.
The thick vertical solid line in the left panel of Fig. \ref{fig:strategy}
indicates the FWHM beam size at the PHI1 channel,
whereas the thin one shows three times the PHI1 beam.
The radio flux density,
of which the corresponding far-infrared intensity is detectable 
in the $5'\times5'$ mapping 
with an overhead-inclusive exposure time of $1$ and $10$ h,
is drawn with the thin and thick horizontal dashed lines, respectively,
in the left panel. 
Similar information is displayed 
in the right panel of Fig. \ref{fig:strategy} for the PPI1 channel. 
The hot spots located in the upper right area of these lines 
are inferred to be accessible  with the PPI or PHI.
Figure \ref{fig:strategy} demonstrates that 
a combination of PPI and PHI is very useful for a wide range 
of hot spots. 
This figure also reveals that these two imagers 
installed in PRIMAger are complementary 
to each other in the sense that the PPI is effective 
for relatively faint sources,
whereas the smaller beam size of the PHI is powerful to 
targets with a small nuclear separation.

\section{Use Cases} 
\label{sec:candidates}
To more specifically demonstrate the ability of PRIMAger 
for the present study, 
this section picks up three hot spots as possible use cases
from the sample of well-studied hot spot \cite{Hardcastle2004,Werner2012,Mack2009,Meisenheimer1997,
Cheung2005,kataoka2005}, 
compiled in Sec. \ref{sec:strategy}. 

\subsection{South Hot Spot of 3C 33} 
\label{sec:3C33}

\begin{figure}[t]
    \centering
    \includegraphics[width=0.7\textwidth]{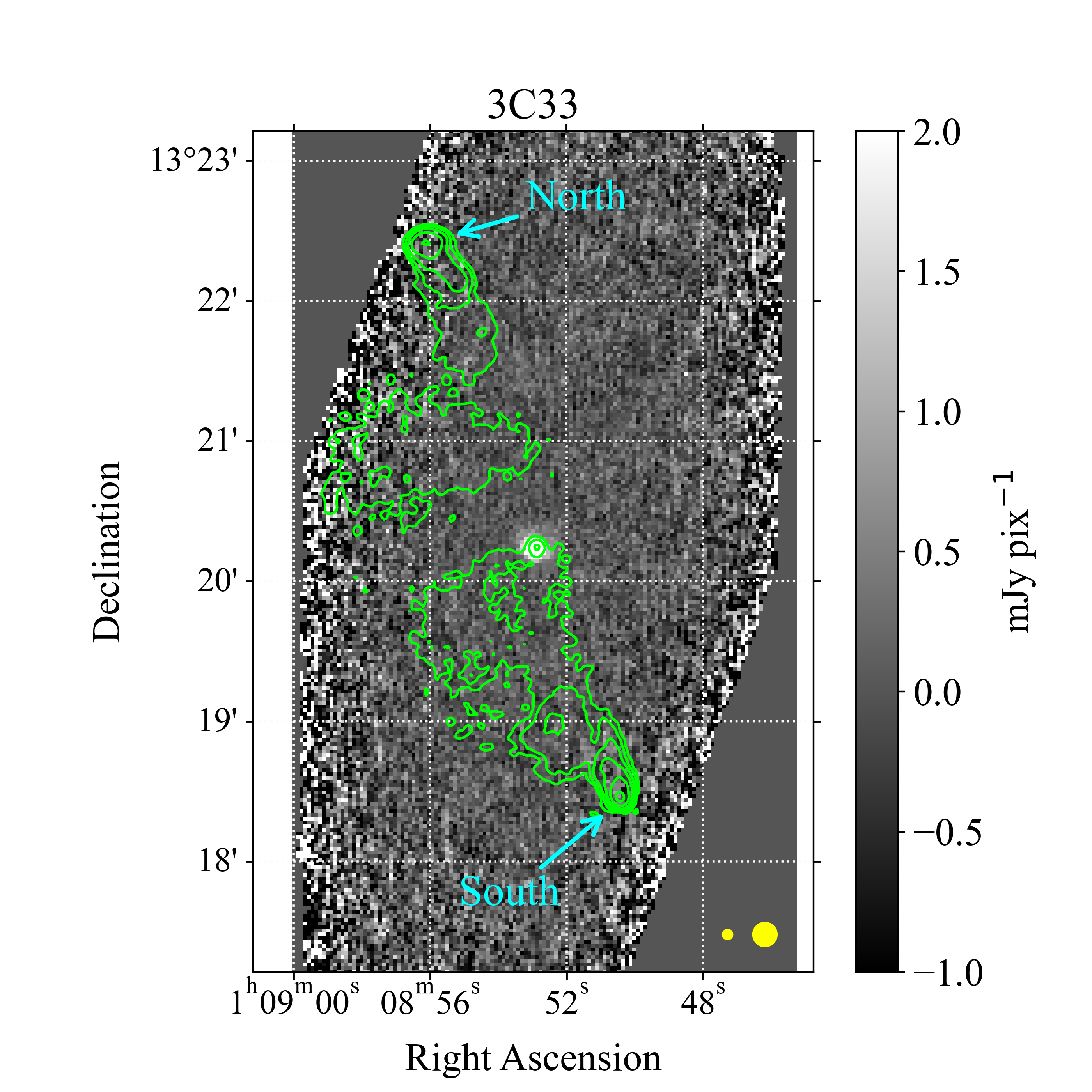}
    \caption{1.5 GHz radio contours \cite{Leahy1991} of the radio galaxy 3C 33, 
    overlaid on the far-infrared image at $\lambda=100$ $\mu$m 
    obtained with the Herschel PACS (ObsID=1342261864). 
    The south and north hot spots of this radio galaxy 
    are indicated with the arrows. 
    The smaller and larger circles in the bottom right display
    the FWHM beam size of the PHI1 ($4.1$ arcsec) and 
    PPI1 ($10.8$ arcsec) channels, respectively.
    The angular distance of $1$ arcsec on the image is converted 
    into the physical distance of $1.16$ kpc at the source frame.}
    \label{fig:3C33}
\end{figure}

The 1.5 GHz radio image  of the radio galaxy 3C 33 \cite{Leahy1991}
is compared in Fig. \ref{fig:3C33}
with the $100$ $\mu$m far-infrared image taken with 
the Photodetector Array Camera and Spectrometer (PACS) 
onboard the Herschel observatory.
Located at the redshift of $z = 0.06$, 
this radio galaxy is known to host two prominent hot spots
as shown with the arrows in Fig. \ref{fig:3C33}.
The PACS image reveals no far-emission associated with the two hot spots.

Among the two hot spots, 
the south one is detected at the optical wavelength range 
with a K-band flux density of $S_\nu = 44 \pm 5$ $\mu$Jy \cite{Meisenheimer1997}.
The south hot spot is plotted with the diamond 
in Fig. \ref{fig:strategy}.
Through a detailed modeling of the radio-to-optical synchrotron spectrum,
the south hot spot is expected to exhibit a high far-to-mid infrared intensity 
at the level of  $\nu S_\nu \gtrsim 10^{10}$ Jy Hz 
in the frequency range of $\nu = 10^{12}$--$10^{13}$ Hz. 
Thus, the south hot spot is detectable 
in the PRIMAger nominal one-square-degree survey (see Fig. \ref{fig:sed}). 
Considering the source size ($\sim 4$ arcmin in length),
the proposed $5'\times5'$ mapping safely covers 
the entire radio structure from the north to south hot spots.
As the angular separation from the nucleus 
($\sim 2$ arcmin) is significantly larger 
than the PRIMAger beam size 
plotted with the circles in Fig. \ref{fig:3C33},
the south hot spot is probably free from the contamination 
from the nuclear emission.
It is important to note that a possible cooling break 
at the frequency of $\nu_{\rm} \sim 3 \times 10^{12}$ Hz 
was discussed in Ref. \citenum{Meisenheimer1997}.
In addition, the spatially averaged magnetic field strength 
was estimated as $B\gtrsim200$ $\mu$G.

Through spatially resolved multifrequency spectroscopy \cite{Kraft2007}, 
which complements the cooling break method enabled 
by the spatially ``integrated'' spectroscopy, 
the compact radio peak at the head of the south hot spot 
is reported to exhibit a similar strength of the magnetic field 
($B\sim 195$ $\mu$G in the S1 region in Ref. \citenum{Kraft2007})
with a slightly higher break frequency ($\nu_{\rm b} \sim 10^{13}$ Hz). 
The positional dependence of the break frequency 
is consistent with the picture of radiative cooling. 
By adding the far-infrared data with PRIMAger, 
a more comprehensive view of the south hot spot of 3C 33 is expected to be derived. 
Thus, this object is a good first-step target 
to validate the power of PRIMAger.

\subsection{South Hot Spot of 3C 173.1} 
\label{sec:3C173.1}
\begin{figure}[t]
    \centering
    \includegraphics[width=0.7\linewidth]{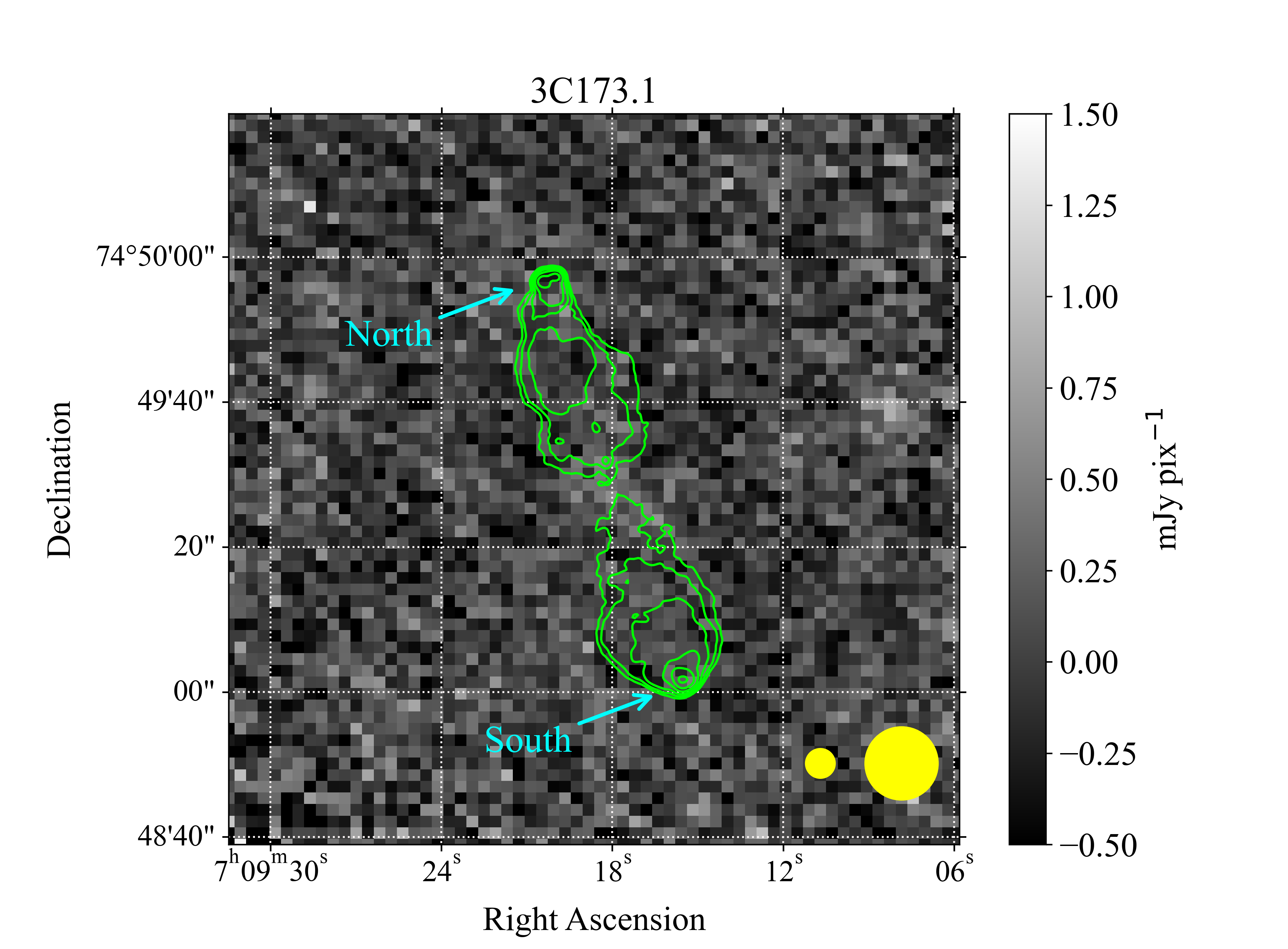}
    \caption{1.5 GHz radio contours 
    of the radio galaxy 3C 173.1 \cite{Leahy1991}, 
    overlaid on the Herschel PACS image 
    at $\lambda = 100$ $\mu$m (ObsID=1342265541). 
    The arrows and circles are written in the same manner as in Fig. \ref{fig:3C33}.
    At the frame of this radio galaxy, the angle scale of $1$ arcsec corresponds 
    to the physical scale of $4.4$ kpc.}
    \label{fig:3C173.1}
\end{figure}

As shown in Fig. \ref{fig:3C173.1}, 
there are two hot spots in the radio galaxy 3C 173.1 ($z = 0.292$).
As the radio flux of the north hot spot is relatively low, 
$S_\nu (5~{\rm GHz})=9$ mJy \cite{Hardcastle2004},
the south one is selected as the main target here. 
The object is plotted with the circle in Fig. \ref{fig:strategy}.

The radio flux and size of the south hot spot are reported 
as $S_\nu (5~{\rm GHz})=33$ mJy and $L=3.6$ kpc (or $0.83$ arcsec)
\cite{Hardcastle2004}, respectively. 
From this object, weak X-ray emission with a 1 keV flux density of 
$S_\nu =0.2$ nJy was detected at the confidence level of $\sim 3 \sigma$.
By reproducing the X-ray flux with a simple SSC model,
Ref. \citenum{Hardcastle2004} suggested that the magnetic field 
in the hot spot is weaker than the equipartition value 
by a factor of $\sim 90$.
Reference \citenum{Werner2012} reconfirmed this result 
by including the mid-infrared data obtained with the {\it Spitzer} observatory,
$S_\nu = 5.1 \pm 0.3$ $\mu$Jy at the wavelength of $\lambda = 3.6$ $\mu$m
($\nu = 8.3 \times 10^{13}$ Hz).
The magnetic field strength of the object is evaluated 
as $B = 1.53$ $\mu$G,
which is  weaker by $2$ orders of magnitude
than the typical value of the well-studied hot spots.
Thus, it is important to measure the magnetic field strength 
in the hot spots through different approaches.

As shown in Fig. \ref{fig:3C173.1},
no significant far-infrared emission has yet been detected 
from the south hot spot of 3C 173.1.
By searching for the cooling break, 
PRIMAger is expected to yield an independent estimate 
for the magnetic field strength 
in this object and to verify the above scenario. 
If the magnetic field in this hot spot is really weak 
as $B=1.53$ $\mu$G \cite{Werner2012},
the object is expected to show practically no spectral break
because its cooling break is inferred to be located in the X-ray range,
i.e., $\nu_{\rm b} = 4.5 \times 10^{17}$ Hz for $L_{\rm cool} = 3.6$ kpc. 
If the magnetic field of the hot spot is, by contrast,
within the typical range 
of the well-studied hot spots, $B=100$--$300$ $\mu$G \cite{kataoka2005}, 
the corresponding cooling break frequency is estimated to be 
within or below the PRIMA frequency range as  
$\nu_{\rm b} = 6.0\times10^{10}$--$1.6\times10^{12}$ Hz. 
Thus, a combination of the PRIMA and radio data is useful 
to specify the cooling break. 

Figure \ref{fig:strategy} suggests that 
by performing the $5'\times5'$ mapping,
the PPI is possible to detect the south hot spot of 3C 173.1 
in a reasonable exposure time of $\sim 3$ h,
whereas the PHI requires more than 10 h for detection. 
However, 
a comparison of the total radio extension of this radio galaxy 
($\sim 1$ arcmin $\times 20$ arcsec) to the instantaneous instrumental field of view 
indicates that the PHI mapping area is possible to be reduced, 
e.g., to $\sim 4$ arcmin $\times2$ arcmin.
The reduction in the mapping area shortens the exposure time 
by a factor of $\sim 3$ into $\gtrsim 3$ h.
This narrower mapping strategy for PHI is commonly applicable 
to sources with a size smaller than $\sim4$ arcmin.

As shown with the circles in Fig. \ref{fig:3C173.1},
the PRIMAger beam size is basically smaller than the angular distance 
between the nucleus and hot spot ($\sim 30$ arcsec). 
However, 
especially at the longer wavelength bands 
(e.g., at the PPI4 channel with the FWHM beam size of $27.6$ arcsec), 
a careful evaluation of the nuclear contamination onto the hot spot 
(e.g., by the beam-convolved image fitting procedure)
is possibly needed to enhance the reliability of the flux measurement.

\subsection{3C 6.1} 
\label{sec:3C6.1}
\begin{figure}[t]
    \centering
    \includegraphics[width=0.7\linewidth]{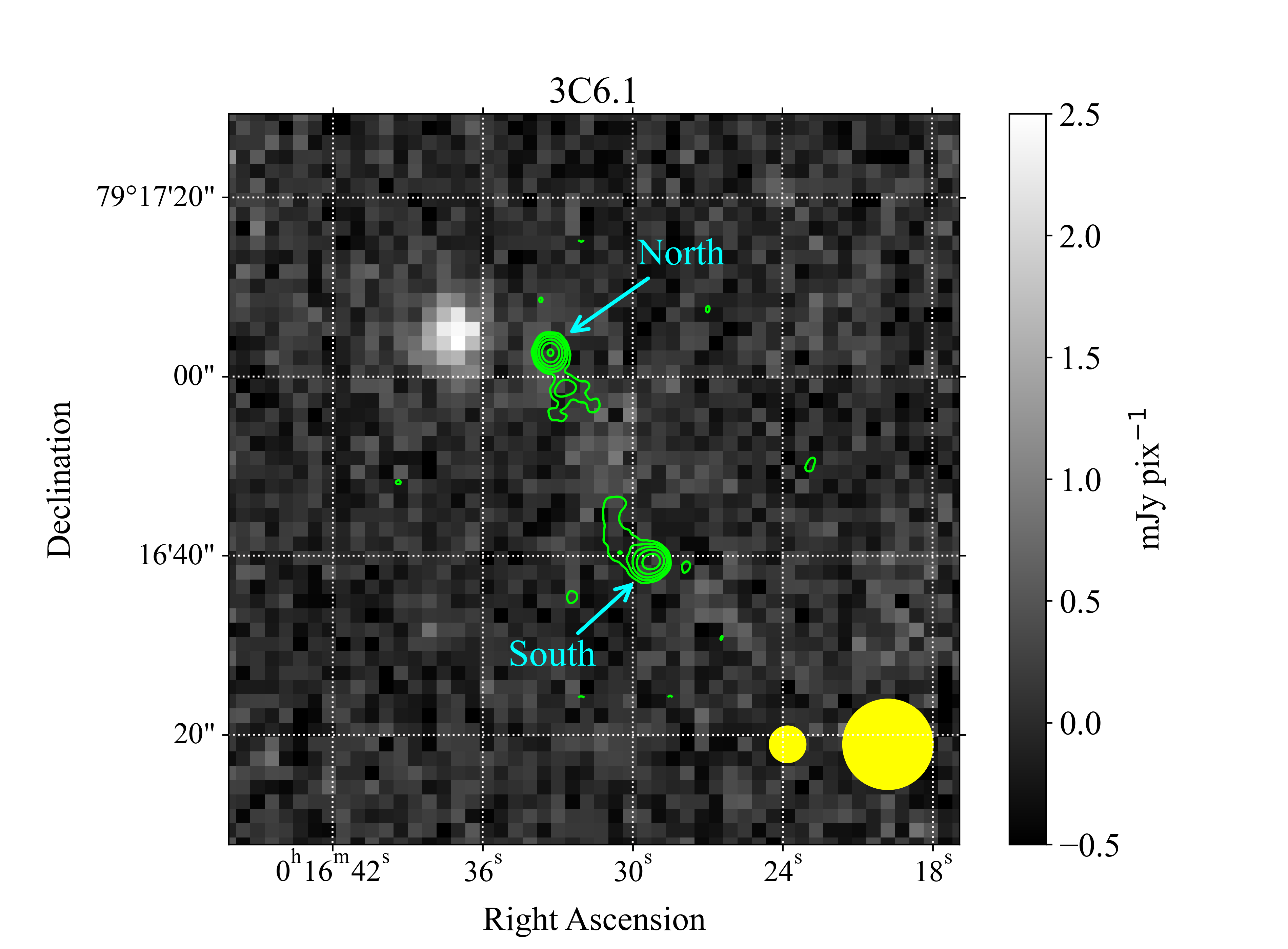}
    \caption{4.9 GHz radio contours 
    of the radio galaxy 3C 6.1 \cite{Neff1995}, 
    superposed onto the Herschel PACS image 
    at $\lambda = 100$ $\mu$m (ObsID=1342262062). 
    The arrows and circles are written in the same manner as of Fig. \ref{fig:3C33}.
    The angle-to-size scaling factor for this object is $7.7$ kpc arcsec$^{-1}$.}
    \label{fig:3C6.1}
\end{figure}

Located at the redshift of $z = 0.8404$,
the radio galaxy 3C 6.1 hosts two X-ray-detected hot spots \cite{Hardcastle2004}, 
the north and south ones.
Figure \ref{fig:3C6.1} displays  $4.9$ GHz radio image of this radio galaxy \cite{Neff1995}, 
superposed on the $100$ $\mu$m Herschel PACS one,
which reveals no significant far-infrared emission from the two hot spots.
Both north and south hot spots exhibit a relatively high radio intensity, 
$S_\nu = 0.34$ and $0.20$ Jy, respectively 
at the frequency of $\nu = 5$ GHz.
Reference \citenum{Werner2012} imposed an upper limit 
on the mid-infrared flux density of the north hot spot
with Spitzer as $S_\nu <1.1$ $\mu$ Jy at $\lambda = 3.6$ $\mu$m,
whereas the south hot spot was reported to be confused in the mid-infrared range. 
Thus, the north hot spot has a higher priority for the PRIMA study.
As pointed out in the box in Fig. \ref{fig:strategy},
the exposure time required to detect the north hot spot 
in the $5'\times5'$ mapping is expected to be significantly 
shorter than 1 h for both PPI and PHI.

When the observed radio-to-X-ray spectral energy distribution 
of the north hot spot, including its mid-infrared upper limit, 
is described with the simple SSC model,
the magnetic field in this object is estimated 
as $B\sim 100$ $\mu$G \cite{Werner2012}. 
The model also predicts a relatively high far-infrared intensity 
as $\nu S_\nu > 10^{10}$ Jy Hz around $\nu = 10^{12}$--$10^{13}$ Hz. 
Considering its source size, $L=2.8$ kpc, 
the north hot spot is expected to show the cooling break 
around $\nu_{\rm b} \sim 2.6 \times 10^{12}$ Hz
at the source rest frame.
With a relatively high redshift of this radio galaxy taken into account,
the break is shifted into $\nu_{\rm b, obs} \sim 1.4 \times 10^{12}$ Hz
at the observers' frame.
Due to a lack of far-infrared data 
(see the PACS image in Fig. \ref{fig:3C6.1}),
the break has not yet been taken into account 
in the previous spectral investigation \cite{Hardcastle2004,Werner2012}.
The PHI,
with a frequency range of $(3.6$--$12.5)\times 10^{12}$ Hz,
is utilized to evaluate the spectral index just 
above the cooling break frequency. 
A combination of the PHI data with the radio one 
enables the measurement of the break frequency. 
The cooling break is expected to reside 
in the PPI spectral coverage.
However, the total angular size this radio galaxy 
($\gtrsim 20$ arcsec as shown in Fig. \ref{fig:3C6.1})
prevents the north hot spot from being resolved with the PPI.
In addition, Fig. \ref{fig:3C6.1} reveals 
a bright far-infrared source 
to the $\sim 10$ arcsec east of this target. 
Therefore, the PHI is regarded as the main instrument 
for the north hot spot of 3C 6.1, 
also from the point of view of the source contamination.

\section{Synergy}  
\label{sec:synergy}
\subsection{Synchrotron Observations} 
\label{sec:synergy_sync}
To systematically apply the cooling-break technique 
to a large number of hot spots, 
a wide spectral coverage is desirable. 
In Fig. \ref{fig:sed}, the frequency ranges of ALMA and JWST (MIRI and NIRCam)
are compared with that of PRIMAger by the arrows. 
A combination of these three facilities enables 
a wide continuous frequency coverage in nearly $4$ orders of magnitude. 

The ALMA data are useful for hot spots which exhibit 
a low-frequency cooling break 
at $\nu_{\rm b} \ll 8 \times 10^{11}$ Hz.
This means that ALMA aims at sources 
with a higher magnetic field and/or a larger physical size 
(see Fig. \ref{fig:break_map}). 
For the PRIMAger targets, 
ALMA is utilized to measure the spectral index below the break 
to validate the standard strong shock (i.e., $\alpha_1 = 0.5$) 
and/or standard cooling break (i.e., $\Delta \alpha = 0.5$) conditions.
For weak-field and/or small-sized (i.e. a high-frequency break) hot spots,
JWST is expected to possess a predominance.
For sources with $\nu_{\rm b} \gg 10^{13}$ Hz,
the degeneracy between the cooling-break and the cut-off frequencies
is expected to be problematic.
This issue is inferred to be settled by the JWST data,
in combination with the PRIMA one.

\subsection{IC X-ray Observations} 
\label{sec:synergy_IC}
Previously, the $S_{\rm X}/S_{\rm R}$ ratio has long been 
widely utilized to evaluate the magnetic field strength 
in the hot spots \cite{Hardcastle2004,kataoka2005}.
Due to limited observational spectral coverage,
the simple PL model is assumed in the SSC calculation 
to reproduce the observed $S_{\rm X}/S_{\rm R}$ ratio.
However, the standard particle-acceleration model
for the hot spots (i.e., the diffusive shock acceleration 
under the continuous energy-injection condition)
predicts a broken PL model (see Fig. \ref{fig:sed}).
Therefore, the $S_{\rm X}/S_{\rm R}$ method 
is known to be subjected to this spectral simplification. 

By incorporating the cooling break $\nu_{\rm b}$ into the SSC model, 
the magnetic-field measurement accuracy 
through the $S_{\rm X}/S_{\rm R}$ ratio is expected 
to be significantly improved. 
Actually, this technique was pioneeringly applied to the hot spot D 
of the prototypal FR-II radio galaxy Cygnus A \cite{Sunada2022}.
In Ref. \citenum{Sunada2022},
the Herschel data played a key role 
in detecting the cooling break of the object at the frequency of 
$\nu_{\rm b} = 2.0^{+1.2}_{-0.8}\times 10^{12}$ Hz
(the corresponding wavelength of $\lambda = 94$--$250$ $\mu$m).
With the observed cooling break frequency 
consistently implemented by adopting the broken PL spectral shape,
the SSC model successfully described 
the multi-wavelength synchrotron and IC spectral energy distribution
of this hot spot.
As a result, 
the magnetic field strength in this object 
was evaluated as $B=120 - 150$ $\mu$G.
This value was found to be smaller by a factor of two
than the previous result obtained by the SSC model
without including the Herschel data\cite{Stawarz2007}.
As a result, an electron dominance of $U_{\rm e}/U_{\rm B} \sim 4$ 
was unveiled in the hot spot D of Cygnus A. 
This strongly suggests that the results from the simple PL-based SSC analysis 
are recommended to be reconsidered with the cooling break taken into account. 

Generally, the signal statistics of IC X-ray emission 
from the hot spots in previous studies were relatively low \cite{Hardcastle2004,kataoka2005}.
Thus, the application of the $S_{\rm X}/S_{\rm R}$ technique 
has been frequently limited by the IC X-ray data. 
In addition, because a higher electron energy density in the hot spot
is expected to yield a higher IC X-ray flux density
for a fixed radio flux density,
the $S_{\rm X}/S_{\rm R}$ result tends to be biased 
to higher $U_{\rm e}/U_{\rm B}$ objects.
Without being affected by such bias, 
the $\nu_{\rm b}$  approach is probably of prominent power 
to magnetically dominated hot spots 
with the $U_{\rm e}/U_{\rm B}$ ratio smaller than unity. 
Actually, by applying this method to the mid-to-far infrared excess
discovered with the WISE and Herschel observatories
from the west hot spot 
of the radio galaxy Pictor A \cite{Isobe2017,Isobe2020,Isobe2023},
the magnetic field strength was suggested to be higher 
by a factor of $\sim 10$ than 
that in the minimum-energy condition \cite{Miley1980},
which is known to be nearly equivalent 
to the $U_{\rm e}$--$U_{\rm B}$ equipartition.
In Refs. \citenum{Isobe2017,Isobe2020,Isobe2023},
this strong magnetic field was successfully attributed 
to field amplification via the post-shock turbulence.
The above consideration indicates that 
cooperation between the two methods covers hot spots 
with a wide range of the $U_{\rm e}/U_{\rm B}$ ratio.

In principle, by combining the $S_{\rm X}/S_{\rm R}$ 
and $\nu_{\rm b}$ techniques,
it is possible to constrain another physical parameter 
additional to the magnetic field.
In Ref. \citenum{Sunada2022}, 
because the reported size of the hot spot D of Cygnus A was found to vary 
by a factor of 2 among publications, 
the authors constrained simultaneously 
the magnetic field $B$ and region size $L_{\rm cool}$.
By contrast, the size of most hot spots picked up in Sec. \ref{sec:strategy} is 
regarded as already known with a relatively good accuracy \cite{Hardcastle2004}.
One of the most important observationally unresolved parameters 
in the $\nu_{\rm b}$ method is the plasma flow velocity $v$ 
in the shock downstream region (see Sec. \ref{sec:prelim_investigatoin}).
The cooling break is determined by the magnetic field and flow velocity,
whereas the $S_{\rm X}/S_{\rm R}$ ratio is 
sensitive to the magnetic field. 
Therefore, by making most of the two observational indicators, 
it is able to solve simultaneously $B$ and $v$.
The measurement of $v$ is expected to shed light 
on one very essential question in modern astrophysics, 
i.e., whether jets from the nucleus of FR-II radio galaxies 
are really relativistic at the position of their hot spots.

\section{Application to the Lobes} 
\label{sec:application}
The $\nu_{\rm b}$ method itself is conceptually applicable to lobes 
of FR-II radio galaxies,
which are filled with the magnetized plasma supplied by the hot spots.
In the case of the lobes, 
because of their large size (e.g., $\gtrsim 100$ kpc) 
and slow plasma flow inside them 
(typically $v<0.1c$ where the hot-spot advance speed is included), 
the break frequency has been typically 
observed in the radio band as $\nu_{\rm b} \ll 100$ GHz 
\cite{Carilli1991,Jamrozy2008,Harwood2013,Harwood2015,Harwood2017}.
Thus, the classical spectral aging technique 
(i.e., investigation into the positional variation of the break frequency)
has widely prevailed in the radio band. 
By contrast, PRIMAger is expected to grasp the higher end 
of its synchrotron spectrum above the cooling break. 

For a lobe to be resolved with PRIMAger
from the nucleus and/or hot spots,
its spatial scale is conservatively requested to be larger than $\gg 30$ arcsec. 
However, when the size and radio flux density of previously well-studied lobes 
associated with FR-II radio galaxies \cite{Croston2005}
are investigated,
only a few objects are anticipated to be detectable with PRIMAger.

As one of the few possible cases,
the observational feasibility with PRIMAger 
for the lobes of the FR-II radio galaxy 3C 452 is discussed here. 
The 1.4 GHz radio flux density integrated over the lobes
is known to be higher by a factor of $\sim5$
than those of the well-studied lobes of 3C radio galaxies 
with a size of $\gg 30$ arcsec \cite{Croston2005}. 
Through the spectral aginganalysis,
the break is suggested to be located in the GHz range \cite{Harwood2017}.
In Ref. \citenum{Isobe2002}, the envelope of the lobes 
was reported to be approximated by a simple ellipse 
with a major and minor radius of $2.46$ and $1.23$ arcmin 
(corresponding to $113$ kpc and $226$ kpc in physical size 
at the source redshift $z=0.0811$).
The high radio intensity and relatively large angular size 
make the lobes an adequate PRIMAger target.
In addition, their total extension is reasonably covered 
with a single $5'\times5'$ mapping proposed in Sec. \ref{sec:strategy}.

By adopting the 5 GHz radio flux, 
$S_\nu (5~{\rm GHz}) = 3.2$ Jy\cite{Laing1980},
the radio surface brightness spatially averaged over the lobe envelope  
is estimated as $f_\nu (5~{\rm GHz}) = 4.0$ MJy sr$^{-1}$.
A simple PL extrapolation of the 5 GHz surface brightness 
with the radio synchrotron spectral index $\alpha = 0.78$ \cite{Laing1983}
indicates the far-infrared surface brightness of the lobes 
as $f_\nu = 0.053$ and $0.025$ MJy str$^{-1}$ 
at the PPI4 ($1.28 \times 10^{12}$ Hz)  
and PPI1 ($\nu = 3.26 \times 10^{12}$ Hz) ranges, respectively.
When the PRIMAger diffuse sensitivity 
in the nominal one-square-degree survey,
0.18 and 0.46 MJy sr$^{-1}$ for the PPI4 and PPI1 channels
as of 2025 February respectively,
is simply scaled by the survey area and total exposure time, 
it is found that an exposure time of 
$t_{5'\times5'} = 0.81$ (PPI4) and $23.7$ (PPI1) hours 
is necessary to detect such 
a low-level far-infrared emission in the $5'\times5'$ mapping.
In a similar manner, the far-infrared surface brightness of the lobes 
in the PHI range is evaluated 
as $f_\nu = 0.023$ MJy sr$^{-1}$ 
at $\nu = 3.57 \times 10^{12}$ Hz ($\lambda=84$ $\mu$m, PHI2) 
and $0.008$ MJy str$^{-1}$ 
at $\nu = 12.5 \times 10^{12}$ Hz ($\lambda=24$ $\mu$m, PHI1).
By considering the PHI nominal survey sensitivity 
at the corresponding frequencies,
0.58 MJy sr$^{-1}$ (PHI2) and 1.64  MJy sr$^{-1}$ (PHI1), 
the required exposure time is computed 
as $t_{5'\times5'} = 43.7$ (PHI2) hours and $2.9 \times 10^3$ (PHI1) hours. 
The above arguments indicate that 
the PPI (especially its longer-wavelength channels) 
is of prime importance for the study of the lobes. 

Finally, as an interesting target, 
the inner lobes of the famous radio galaxy Centaurs A are discussed. 
Even though the radio galaxy is classified as an FR-I radio source,
it exhibits a hierarchical lobe structure, i.e.,  
from the inner to outer ones.
These lobes are widely regarded 
as one of the most promising candidates 
for ultra-high-energy cosmic-ray accelerators 
\cite{Aab2018,Matthews2018}. 
With Spitzer, mid-infrared emission was detected 
from bright radio regions inside 
the north inner lobe of this radio galaxy,
at the $24$ $\mu$m peak surface-brightness level 
of $\gtrsim 30$ MJy sr$^{-1}$ \cite{Hardcastle2006}. 
The same regions are reported to be detected 
at the wavelength of $\lambda = 870$ $\mu$m ($\nu = 345$ GHz) \cite{Weiss2008}
by the Large Apex Bolometer Camera operated 
at the Atacama Pathfinder Experiment telescope. 
By combining these two results, 
the bright regions inside the north inner lobe are suggested to exhibit 
a spectral break in the range of $(0.5$--$3)\times10^{12}$ Hz,
with a non-standard break condition ($\Delta \alpha \sim 0.6$) \cite{Weiss2008}.
In addition, Ref. \citenum{Parkin2012} reported 
a far-infrared detection of the north and south inner lobes of this radio galaxy 
at the wavelength of $\lambda = 350$ and $500$ $\mu$m
($\nu =8.6\times10^{11}$ and $6.0\times10^{11}$ Hz, respectively)
with the Spectral and Photometric Imaging Receiver (SPIRE)
onboard the Herschel observatory, 
although no detailed photometric analysis was performed. 
PRIMAger is expected to fill the frequency gap 
between Spitzer and the Herschel SPIRE,
where the cooling break is anticipated,
and to enable a precise measurement of the magnetic field strength 
not only in the north inner lobes but also 
in the south one of Centaurus A.
The PRIMAger observation of this object will be useful 
to examine differences in the physical condition 
between FR-I and FR-II jets.

\section{Summary}
It is discussed that PRIMAger is a useful probe 
to investigate the particle acceleration phenomena
associated with the jets from active galactic nuclei.
The present study focuses on the hot spots of radio galaxies
as one of the promising sites of diffusive shock acceleration.
To determine the magnetic field in the hot spots,
which is the most important parameter 
to specify the acceleration condition,
the synchrotron spectral feature called the cooling break is 
indicated to be useful
because it is sensitive to the magnetic field strength.
From the typical parameters of nearby well-studied hot spots, 
the frequency of the cooling break is estimated 
to reside in or below the PRIMAger frequency range. 
A realistic observational strategy with PRIMAger 
for the hot spots is described.
It is shown that the systematic magnetic-field evaluation 
for the hot spots though the cooling break 
from PRIMAger observations is quite feasible.
To more specifically demonstrate the PRIMAger potential, 
observations of three hot spots are presented as a use case.
The cooling-break method is advantageous from the viewpoint that 
the magnetic field is measurable from the synchrotron spectrum 
in the submillimeter, far-infrared and mid-infrared ranges alone.
In addition, the synergy with IC X-ray observations is expected 
to enhance the scientific outcome,
by enabling the precise measurement of the magnetic field 
for a wide range of hot spots, 
from electron-dominant to magnetic-dominant ones. 

\subsection*{Disclosures}
No possible conflict of interest is expected. 

\subsection*{Code, Data, and Materials Availability} 
The radio images of the candidate objects shown in Sec. \ref{sec:candidates}
are obtained from the NASA/IPAC Extragalactic Database:
\linkable{https://ned.ipac.caltech.edu/}
The Herschel data utilized in the present paper 
(ObsID = 1342261864, 1342265541 and 1342262062) 
are electrically available from the Herschel Science Archive
at \linkable{http://archives.esac.esa.int/hsa/whsa/}

\subsection*{Acknowledgments}
This research was supported by the JSPS KAKENHI 
(Grant Nos. 21K03635, 23H00130, 23H00134, 23H05441, 
21H04496, 22H00157, and 23K17695).

\appendix    


\bibliography{report.bib}   
\bibliographystyle{spiejour}   


\vspace{2ex}\noindent\textbf{N. Isobe} is an assistant professor 
at the Institute of Space and Astronautical Science (ISAS),
the Japan Aerospace Exploration Agency (JAXA). 
He received his master's and doctor's degrees 
from the Faculty of Science at the University of Tokyo in 1999 and 2001, 
respectively.
He started his research career in the field of X-ray astronomy
and contributed to the development and operation 
of several Japanese X-ray observatories, 
including {\it ASCA}, {\it Suzaku} and {\it MAXI}.
He moved to the field of infrared astronomy 
to join the development of the SPICA mission.
Currently, he is one of the steering members of 
the next-generation Japanese infrared astrometry mission JASMINE.  
Throughout his research career, 
he has been interested in high-energy phenomena associated 
with relativistic jets emanating from supermassive black holes 
located at the center of galaxies.

\vspace{1ex}
\noindent Biographies and photographs of the other authors are not available.

\listoffigures
\listoftables

\end{spacing}
\end{document}